\documentclass[9pt,twocolumn,twoside]{opticajnl}
\journal{opticajournal} 

\usepackage{pgf}
\usepackage{lineno}
\usepackage{graphicx}
\usepackage{bm}
\usepackage[normalem]{ulem}
\linenumbers 

\title{Increasing the secret key rate of satellite-to-ground entanglement-based QKD assisted by adaptive optics}

\author[1,2,*]{Valentina Marulanda Acosta}
\author[3,4]{Daniele Dequal}
\author[2]{Matteo Schiavon}
\author[1]{Aurélie Montmerle-Bonnefois}
\author[5]{Caroline B. Lim}
\author[1]{Jean-Marc Conan}
\author[2]{Eleni Diamanti}

\affil[1]{ONERA, DOTA, Paris Saclay University, F-92322 Châtillon, France}
\affil[2]{Sorbonne Université, CNRS, LIP6, F-75005 Paris, France}
\affil[3]{Telecommunication and Navigation Division, Agenzia Spaziale Italiana, Matera, Italy}
\affil[4]{Connectivity and Secure Communication Directorate, European Space Agency, Noordwijk, Netherlands}
\affil[5]{LNE-SYRTE, Observatoire de Paris, Université PSL, CNRS, Sorbonne Université, LNE, F-75014 Paris, France}

\affil[*]{valentina.marulanda-acosta@lip6.fr}

\begin{abstract}
Future quantum networks will be composed of both terrestrial links for metropolitan and continent-scale connections and space-based links for global coverage and infrastructure resilience. However, the propagation of quantum signals through the atmosphere is severely impacted by the effects of turbulence. This is even more the case for entanglement-based quantum communication protocols 
requiring two free-space channels to be considered simultaneously. In this work, we assess the advantage of turbulence mitigation by adaptive optics, in particular during daytime link operation, so as to increase the coupling of the received signal into an optical fiber. We show in particular that this improves the performance of entanglement-based quantum key distribution by up to a few hundred bits per second when compared with the uncorrected scenario.
\end{abstract}

\setboolean{displaycopyright}{false} 

\begin{document}
\nolinenumbers
\maketitle

\section*{Introduction}

Quantum key distribution (QKD) allows for the exchange of information-theoretically secure cryptographic keys between two parties by exploiting the properties of quantum physics. Optical fiber implementations of QKD protocols have progressed tremendously the past years, reaching distances of up to a thousand kilometers for configurations with an intermediary untrusted node~\cite{PhysRevLett.130.210801}, albeit at very low secret key rates. For intercontinental quantum communication at practical rates, satellite-to-ground architectures are an attractive solution and are also compatible with untrusted node schemes.

More specifically, entanglement-based QKD based on the BBM92 protocol~\cite{PhysRevLett.68.557} allows for two distant optical ground stations to share a secret key through the emission of entangled pairs of photons by a source onboard a satellite visible by both stations. A milestone experimental demonstration of such a scheme with a low earth orbit (LEO) satellite was performed with the Micius satellite~\cite{Yin2020}. In that work, the key exchange was performed between two stations separated by 1120~km, at nighttime and without fiber coupling, and the obtained key rate was 0.43~bit/s and 0.12~bit/s in the asymptotic and finite-size regime, respectively.

Coupling the received light into a single mode fiber (SMF) has the advantage of the fiber acting as a rather strong spatial filter to reduce background noise, especially at daytime, as well as permitting the use of highly efficient superconducting nanowire single photon detectors (SNSPD). Atmospheric propagation, however, results in optical aberrations that severely limit efficient coupling into an SMF. This is particularly significant during the more severe turbulent conditions occurring during the day. To mitigate this effect, adaptive optics (AO), a correction method that reduces wavefront aberrations, now commonly used in free-space optical communications~\cite{tyson2022principles}, is an attractive solution that can be exploited for satellite-based quantum communication as well ~\cite{ecker23,acosta2021analysis}.

In this work, we perform an in-depth study of an entanglement-based configuration and demonstrate the performance improvement of daytime satellite-to-ground BBM92 QKD with fiber coupling and partial turbulence correction by AO.

\section{Theoretical model}\label{channel}
\subsection*{Channel model}

In order to assess the attenuation affecting the quantum signals when propagating through an atmospheric channel, several effects need to be considered, namely the pointing error, the atmospheric turbulence, the geometrical losses and the coupling into an SMF. We stress again that the latter enables the use of SNSPDs while also reducing the background noise via spatial filtering.

We start our analysis by considering the geometrical losses affecting the satellite-to-ground link. As most of the light propagation occurs in vacuum, and since the beam size has already been enlarged by diffraction before entering the atmosphere, the effect of turbulence on the beam size and shape is negligible compared to diffraction. This allows us to estimate the beam waist on ground as $w=\theta_d R$, where $\theta_d$ is the beam divergence, given by the size and quality of the transmitting telescope, and $R$ is the instantaneous satellite-to-ground distance.

After calculating the beam waist on ground, we calculate its displacement with respect to the receiver center (\emph{i.e.}, the beam wandering) and the wavefront distortion. We assume that the former is dominated by satellite pointing jitter, while the latter is induced by atmospheric turbulence. These two effects are therefore assumed to be statistically independent and are treated separately in our analysis.

We quantify the beam wandering effects through the probability distribution of the light collected into the receiver telescope, denoted hereafter $P_{BW}$. In particular, following~\cite{Vasylyev2012}, we assume that the distance of the beam from the center of the receiving aperture follows a Weibull distribution from which one can deduce the distribution $P_{BW}$. For this, we take into account the beam waist on the ground for every given elevation, as well as the pointing jitter standard deviation $\theta_p$.

The effect of turbulence-induced wavefront distortion on fiber coupling is quantified with $P_{AO}$ the associated probability distribution of the SMF coupling efficiency, which is defined as the square norm of the overlap integral between the incoming wavefront and the SMF mode. To estimate this effect, we model turbulence through the altitude profile of the the refractive index structure constant, $C_n^2$, which describes the distribution in altitude of the turbulence strength. For a given line of sight one can also characterize turbulence conditions by the following set of integrated parameters: the Fried parameter $r_0$ that characterizes the overall turbulence strength, and the isoplanatic angle $\theta_0$ that represents the angular evolution of the wavefront~\cite{hardy1998adaptive,tyson2022principles}. 

In our study, similarly to the analysis in~\cite{acosta2021analysis}, we consider daytime profiles constructed from $C_n^2$ high altitude data at an astronomical site in Paranal~\cite{osborn_optical_2018} and low altitude measurements taken at the Canary Islands~\cite{sprung_characterization_2013}. Note that this construction is based on measurements coming from astronomical sites because they are significantly more abundant than data from more urban settings. We then compute the values of the Fried parameter and the isoplanatic angle for all the available data, and we derive the probability distributions for both. From this, we select a profile that represents turbulence conditions that are more severe than 75\% of the database values~\cite{vedrenne_performance_2021}, which corresponds to $r_0$  = 10.6~cm and $\theta_0$ = 25.8$~\mu$rad at the zenith and at a 1.55 $\mu$m wavelength.

We recall here that optical aberrations induced by turbulence may significantly hinder the coupling into an SMF. We therefore consider the use of AO which provides a real time partial correction of these phase aberrations. Such a system constitutes a feedback loop in which: a wavefront sensor (WFS) measures the aberrations of the incoming optical beam; based on these measurements, a real time computer (RTC) computes the appropriate commands so as to drive a deformable mirror (DM) that performs the optical correction of the phase ahead of the injection into the SMF, hence improving the phase matching with the fiber mode.

A common way to model the phase aberrations and their correction by AO is through the decomposition into Zernike polynomials~\cite{noll_zernike_1976}. Each polynomial describes an optical aberration in the receiver telescope aperture. Polynomials can be grouped by radial orders. The number of phase modes $n_{mod}$ from the first radial order (tip and tilt) to the radial order $n_r$ is given by: $n_{mod} = n_r (n_r + 3)/2$. In our analysis the incoming turbulent aberrations are described on a wide space covering 40 radial order (corresponding to 860 Zernike modes), while we perform a parametric study considering AO systems correcting a phase subspace corresponding respectively to 1, 5, 10, 15 or 20 radial orders. In all cases we assume a standard two frame delay AO loop with a temporal frequency set to 5~kHz.

We estimate the effect of such correction schemes on SMF coupling with a pseudo-analytic Monte-Carlo based simulation tool~\cite{Vedrenne2016,canuet_statistical_2018,acosta2021analysis}, which takes into account the previously constructed turbulence profile, the correction capability of the AO system and several elevations associated to a given satellite pass. 

More specifically, we first draw random occurrences of the complex amplitude of the signals. Phase effects are modelled through a detailed error budget that for each scenario estimates the residual phase variance left after AO correction, taking into account three major error sources~\cite{acosta2021analysis}. The first source, called fitting error, is related to the limited correction capacity of the deformable mirror, characterized here by a correction subspace with a limited number of radial orders. The second term is the aliasing error that alters the wavefront measurements and is due to the limited resolution of the sensor. The final contributor is the temporal error induced by the finite time response of the loop and by the temporal dynamics of the turbulence. For each one of the complex amplitude occurrences, we estimate the overlap integral with the optical mode of the SMF, resulting in a coupling efficiency value. After a sufficient number of occurrences (in the order of ten thousand), we derive a probability distribution of the coupling efficiency, $P_{AO}$.

Bringing together all the effects discussed above, we find that at each point of the trajectory, the probability distribution of the transmission efficiency $\tau$ of the quantum channel is given by~\cite{acosta2021analysis}:
\begin{equation}
  PDTE(\tau) = \int_{0}^{\infty}P_{BW}(x)P_{AO}(\tau/x)\frac{1}{|x|}dx,
\end{equation}
where $P_{BW}(x)$ and $P_{AO}(x)$ denote, as we saw earlier, respectively, the probability distribution of the transmission taking into account the beam wandering due to the pointing error and of the coupling efficiency taking into account turbulence and AO effects.

\subsection*{BBM92 key rate estimation}\label{keyrate}

Having established a theoretical channel model relevant for our configuration, we now describe the entanglement-based QKD protocol and corresponding secret key rate expressions that we use as a benchmark for our analysis.

In the asymptotic regime, \emph{i.e.}, assuming that an infinite amount of photon pairs were emitted and received, the secret key rate for the BBM92 protocol can be calculated as~\cite{ma2007quantum}:
\begin{equation}
  K_A = \frac{1}{2}R_c\left(1-f_{ec}h_2(e)-h_2(e)\right),
  \label{asymptotic}
\end{equation}
where $h_2$ corresponds to the binary entropy function, $f_{ec}$ is the efficiency of the error correction code used for basis reconciliation, $e$ is the quantum bit error rate (QBER) and $R_c$ is the coincidence rate at which photons are detected by both receiver stations, called Alice and Bob. The latter is a function of the conditional probability of an erroneous detection given that no pairs were emitted ($Y_0$), the probability of a coincidental detection given that a pair was emitted ($Y_1$) as well as of the probabilities of a pair being emitted or not ($(1-p_0)$ and $p_0$ respectively), as follows:
\begin{equation}
  R_c = \frac{1}{\Delta t}\left(p_0Y_0+(1-p_0)Y_1\right).
  \label{coincidences}
\end{equation}
In the above equation, $\Delta_t$ is the time window in which detection events are possible. $Y_0$ is obtained assuming the background noise follows a Poisson distribution of mean $d_{a(b)}$. In this case, since we are considering a daytime scenario, the main contributor to detection events in the absence of emission is the sky radiance captured within the fiber's field of view~\cite{PhysRevApplied.16.014067}. The overall probability of a coincidence, $Y_1$, depends on the probabilities of a detection by Alice or Bob, $y_{1a(b)}$, and on the efficiency of each channel, $\eta_{a(b)}$. This efficiency depends on the transmission efficiency of each channel, $\tau_{a(b)}$, and the efficiency of the detectors, $\eta_{dA(dB)}$, such that $\eta_{a(b)}$ = $\tau_{a(b)}\eta_{dA(dB)}$. Here, $\tau_{a(b)}$ correspond to the probability distributions modelled in the previous section.
The probability of measuring a coincidence is then given by:
\begin{equation}
  Y_1 = [y_{0a}+\eta_ay_{1a}]\cdot[y_{0b}+\eta_by_{1b}].
  \label{y1}
\end{equation}
We see that $Y_1$ includes contributions from the false coincidence detections due to dark counts, $y_{0a(b)} = 1- exp(-d_{a(b)}\Delta_t)$, and the true coincidence detections due to a signal being received, $\eta_{a(b)}y_{1a(b)}$, with $y_{1a(b)} = 1- y_{0a(b)}$.

Last, the QBER, $e$, is estimated from the aforementioned $Y_{1(0)}$ and $p_0$ probabilities, as well as the probability of an erroneous detection given that a pair was emitted, $e_1$, detections due to background noise, $e_0$, and detections due to photons falling into the visibility field of the wrong detector, $e_d$.

This allows us to calculate the asymptotic secret key rate. However, in realistic scenarios, such as in the case of a LEO satellite that only remains visible for a couple of minutes for a given passage severely limiting the amount of quantum exchanges that can be made, it is imperative to consider finite-size effects. In this case, the secret key rate is given by~\cite{Yin2020}:
\begin{align}
  \begin{split}
    K_{FS} = \frac{1}{T_v}\left[C_T-C_Th_2\left(e+\sqrt{\frac{(C_T+1)\log_2(1/\varepsilon_{\text{sec}})}{4C_T^2}}\right)\right.\\
    \left.-C_Tf_{ec}h_2(e)-\log_2\left(\frac{2}{\varepsilon_{\text{corr}}\varepsilon_{\text{sec}}^2}\right)\right].
    \label{finite-size}
  \end{split}
\end{align}
Here, we estimate the finite-size key rate by calculating the total number of coincidence counts, $C_T$, that are measured during the visibility time of the satellite, $T_v$, for a single pass. We additionally consider two relevant parameters, namely $\varepsilon_{\text{sec}}$ and $\varepsilon_{\text{corr}}$, which correspond to the secrecy and correctness of the final key, respectively~\cite{Tomamichel2012}.

\section{Methodology and results}
\subsection*{Simulation parameters}
In the scenario under study, the entanglement-based BBM92 QKD protocol is performed with a source onboard a satellite following the same trajectory as the Micius satellite~\cite{Yin2020}, that is a sun-synchronous orbit at a 500 km altitude. We assume that entangled photons are generated at a rate of $\mu =  11.4 \cdot 10^{6}$~pairs/s, which corresponds to state-of-the-art entangled photon sources at telecom wavelength developed for terrestrial applications~\cite{PhysRevApplied.18.024059}. Motivated by realistic configurations of communication links at a national and European level, we consider first the case where entangled photons are sent to two optical ground stations (OGS) in the two French cities of Paris and Nice, separated by about 686~km on the ground, and then the case of a link between Nice and the Italian city of Matera, both of which are located in proximity to an OGS and which are separated by a distance of approximately 841~km.

The sky background radiance considered in order to calculate the dark count rate is estimated with the help of the LOWTRAN atmospheric tool~\cite{Kneizys1988}, and the absorption coefficient is computed with the MODTRAN~\cite{modtran} atmospheric tool, both considering \textit{clear sky} conditions with a visibility of 23~km and a solar angle of 45$^\circ$. 

\begin{figure}[H]
  \includegraphics[width=1.1\linewidth]{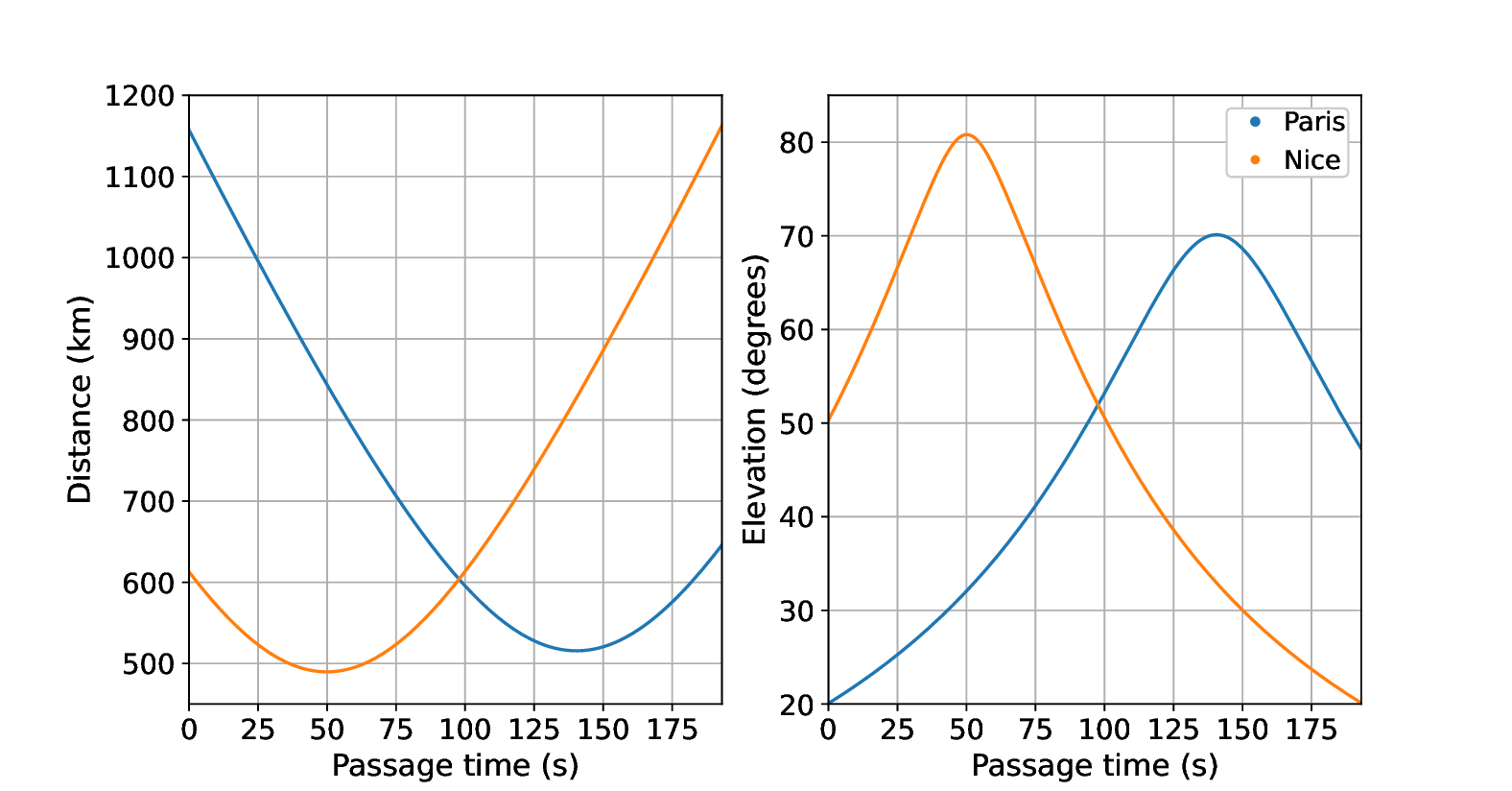}
  \caption{Distance (left) and elevation (right) from the satellite to the Paris and Nice optical ground stations.}
  \label{fig:trajectory}
\end{figure}

The distance from the satellite to each of the two considered locations in Paris and Nice for one satellite pass is shown in Fig.~\ref{fig:trajectory}. We consider only the part of the trajectory where both ground stations observe the satellite
at an elevation of 20$^\circ$ or more, which is compatible with modern advanced adaptive optics and tracking systems.

Since the distance from each station to the satellite is different, the losses experienced on each channel differ as well. Therefore, we divide the trajectory in short time intervals (of around one second) and for each interval we estimate the probability distribution of the transmission efficiency ($PDTE$) for the satellite-to-Paris and satellite-to-Nice channels, using the methods described in Section~\ref{channel}. Based on this, we can compute the average count rate and the average QBER at different points of the orbit. From these, we then derive the secret key rate estimations as per Eqs.~(\ref{asymptotic}) and (\ref{finite-size}). 

Table~\ref{tab:params} summarizes the values of the main parameters used in our simulation, corresponding to state-of-the-art systems and implementations.

\begin{table}[htbp]
  \centering
  \begin{tabular}{|c|c|c|}
    \hline
    \bm{{$\eta_{dA(dB)}$}} = 0.85         & \bm{{$d_{a(b)}$}} = 4.2 $\cdot10^4$ counts/s &\bm{{$\varepsilon_{\text{corr}}$}} = 10$^{-10}$  \\
    \hline
    \bm{$r_0$} = 10.6 cm           & \bm{{$f_{ec}$}} = 1.16                       & \bm{{$\mu$}} = 11.4$\cdot10^6$ pairs/s      \\
    \hline
    \bm{{$D_{Rx}$}} = 1.5 m        & \bm{$\theta_0$} = 25.8 $\mu$rad              & \bm{{$\theta_d$}} = 10 $\mu$rad        \\
    \hline
    \bm{{$\theta_p$}} = 1 $\mu$rad & \bm{{$\Delta t$}} = 500 ps                   &       \bm{{$\varepsilon_{\text{sec}}$}} = 10$^{-10}$                                \\
    \hline

  \end{tabular}
  \caption{Values of the simulation parameters. $\eta_{dA(dB)}$: Detection efficiency, $d_{a(b)}$: Dark count rate, $\mu$: Pair generation rate, $r_0$: Fried parameter at 90$^\circ$ elevation, $f_{ec}$: Error correcting code efficiency, $D_{Rx}$: Receiver diameter, $\theta_0$: Isoplanatic angle at 90$^\circ$ elevation, $\theta_d$: Divergence, $\theta_p$: Pointing error, $\Delta t$: Detection window, $\varepsilon_{\text{sec}}$: Security parameter, $\varepsilon_{\text{corr}}$: Correctness parameter.}
  \label{tab:params}
\end{table}

\subsection*{Results}

\begin{figure}[H]
  \includegraphics[width=0.9\linewidth]{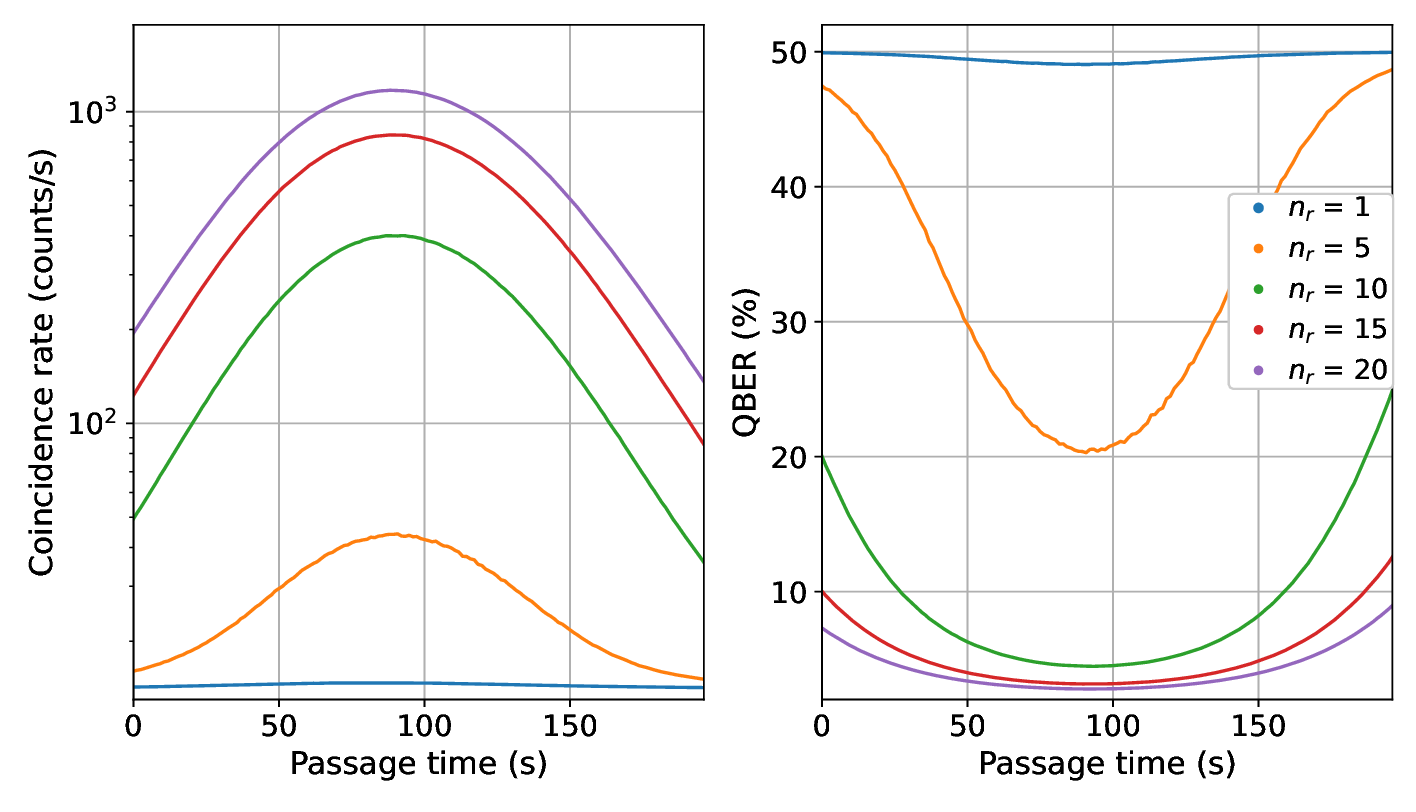}
  \caption{Coincidence rate (left) and QBER (right) of a Paris-Nice link for different AO correction orders ($n_r$)}.
  \label{fig:Coincidence}
\end{figure}

Figure~\ref{fig:Coincidence} shows the coincidence count rate and the QBER calculated for our scenario for different degrees of adaptive optics correction. If we consider no AO correction beyond the basic tip-tilt compensation ($n_r$ = 1) which is in any case mandatory for tracking, the coincidence rate is very low, and the error rate stays solidly
around 50 \% which will completely hinder the establishment of a secret key. A slightly more sophisticated system
capable of correcting up to 5 radial orders already results in significantly improved coincidence rates even if the error rate remains elevated.

\begin{figure}[H]
  \includegraphics[width=0.9\linewidth]{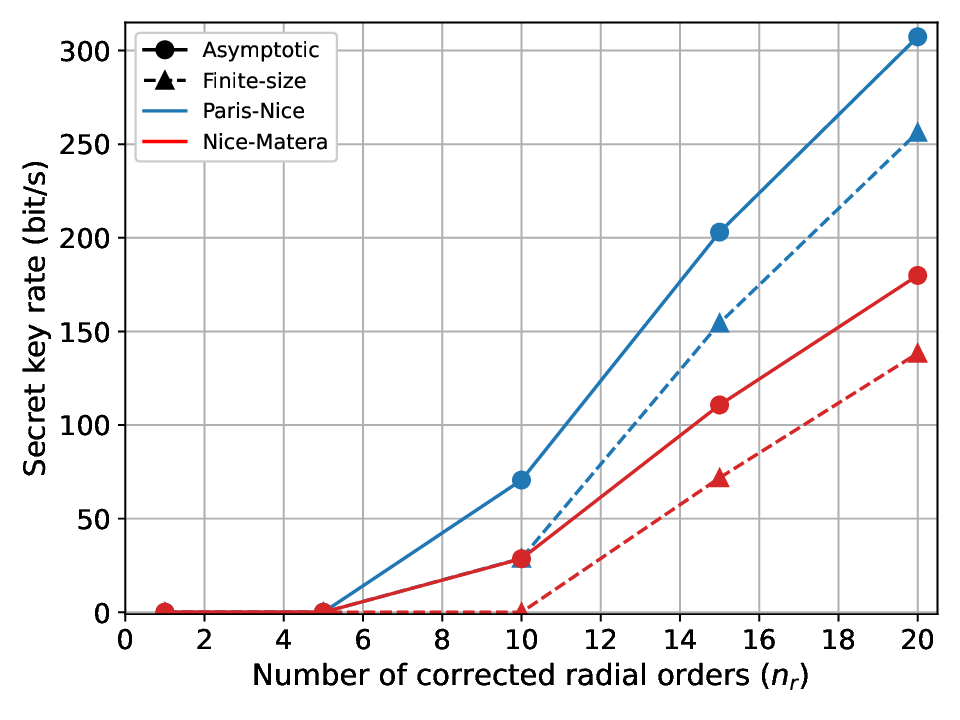}
  \caption{Secret key rate estimation for the Paris-Nice and Nice-Matera links in the asymptotic and finite-size regimes and for different degrees of AO correction.}
  \label{fig:SKR}
\end{figure}

We can then calculate the resulting secret key rate corresponding to the coincidence rate $R_c$ and QBER estimated above in both the asymptotic and finite-size regime, as shown in Fig.~\ref{fig:SKR} for one satellite pass and different levels of AO correction. In this figure, we also show the corresponding results for the alternative scenario of a link between optical ground stations in Nice and Matera using the Micius satellite. This is an interesting situation in the context of a future European quantum communication infrastructure.

As could be predicted from the 40 to 50\% error rates, the systems with no AO or low complexity AO correction are not able to produce any key at all. More sophisticated systems ($n_r$ = 10 and above) show improvement of the average key rate achieved when increasing the correction level. A system correcting 15 radial orders or more would be necessary in order to obtain a key for both scenarios, particularly in the more realistic finite-size estimation.

\section*{Conclusion}

Our simulations analyse the implementation of the entanglement-based BBM92 QKD protocol over space-based links, during daytime and assisted by an adaptive optics system in order to improve coupling into a single-mode fiber. Our results illustrate that AO systems of increasing complexity significantly improve the coincidence count rate and reduce the error rate, resulting in increased key rates, including in the finite-size regime. Somewhat complex but still realistic AO systems correcting 15 radial orders or more would permit to obtain key rates of up to a few hundred bits per second in realistic configurations.

\section{Corresponding author}

Valentina Marulanda Acosta (valentina.marulanda-acosta@lip6.fr)

\begin{backmatter}

\bmsection{Acknowledgments} We acknowledge financial support from the European Union’s Horizon Europe research and innovation programme under the project QSNP, Grant No.~101114043, and the project QUDICE, Grant No.~101082596, the French National Research Agency (ANR) through the project SoLuQS, and the PEPR integrated project QCommTestbed, ANR-22-PETQ-0011, part of Plan France 2030.

\bmsection{Disclosures} The authors declare no conflicts of interest.

\bmsection{Data Availability Statement}
Data underlying the results presented in this paper are not publicly available at this time but may be obtained from the authors upon reasonable request.
\end{backmatter}

\bibliography{main}

\begin{thebibliography}{10}
\newcommand{\enquote}[1]{``#1''}

\bibitem{PhysRevLett.130.210801}
Y.~Liu, W.-J. Zhang, C.~Jiang \emph{et~al.}, \enquote{Experimental {{Twin-Field Quantum Key Distribution}} over 1000 km {{Fiber Distance}},} {\protect\JournalTitle{Physical Review Letters}} \textbf{130} (2023).

\bibitem{PhysRevLett.68.557}
C.~H. Bennett, G.~Brassard, and N.~D. Mermin, \enquote{Quantum cryptography without {{Bell}}'s theorem,} {\protect\JournalTitle{Physical Review Letters}} \textbf{68} (1992).

\bibitem{Yin2020}
J.~Yin, Y.-H. Li, S.-K. Liao \emph{et~al.}, \enquote{Entanglement-based secure quantum cryptography over 1,120 kilometres,} {\protect\JournalTitle{Nature}} \textbf{582} (2020).

\bibitem{tyson2022principles}
R.~K. Tyson and B.~W. Frazier, \emph{Principles of {{Adaptive Optics}}} ({CRC Press}, 2022), 5th ed.

\bibitem{ecker23}
S.~Ecker, J.~Pseiner, J.~Piris \emph{et~al.}, \enquote{Advances in entanglement-based {{QKD}} for space applications,} {\protect\JournalTitle{Proc. SPIE, International Conference on Space Optics — ICSO 2022}} \textbf{12777} (2023).

\bibitem{acosta2021analysis}
V.~Marulanda~Acosta, D.~Dequal, M.~Schiavon \emph{et~al.}, \enquote{Analysis of satellite-to-ground quantum key distribution with adaptive optics,} {\protect\JournalTitle{New Journal of Physics}} \textbf{26} (2024).

\bibitem{Vasylyev2012}
D.~{\relax Yu}. Vasylyev, A.~A. Semenov, and W.~Vogel, \enquote{Toward {{Global Quantum Communication}}: {{Beam Wandering Preserves Nonclassicality}},} {\protect\JournalTitle{Physical Review Letters}} \textbf{108} (2012).

\bibitem{hardy1998adaptive}
J.~W. Hardy, \emph{Adaptive Optics for Astronomical Telescopes}, no.~16 in Oxford Series in Optical and Imaging Sciences ({Oxford University Press}, 1998).

\bibitem{osborn_optical_2018}
J.~Osborn, R.~W. Wilson, M.~Sarazin \emph{et~al.}, \enquote{Optical turbulence profiling with {{Stereo-SCIDAR}} for {{VLT}} and {{ELT}},} {\protect\JournalTitle{Monthly Notices of the Royal Astronomical Society}} \textbf{478} (2018).

\bibitem{sprung_characterization_2013}
D.~Sprung and E.~Sucher, \enquote{Characterization of optical turbulence at the solar observatory at the {{Mount Teide}}, {{Tenerife}},} {\protect\JournalTitle{Proc. SPIE, Remote Sensing of Clouds and the Atmosphere XVIII; and Optics in Atmospheric Propagation and Adaptive Systems XVI}} \textbf{8890} (2013).

\bibitem{vedrenne_performance_2021}
N.~V{\'e}drenne, C.~Petit, A.~{Montmerle-Bonnefois} \emph{et~al.}, \enquote{Performance analysis of an adaptive optics based optical feeder link ground station,} {\protect\JournalTitle{Proc. SPIE, International {{Conference}} on {{Space Optics}} --- {{ICSO}} 2020}} \textbf{11852} (2021).

\bibitem{noll_zernike_1976}
R.~J. Noll, \enquote{Zernike polynomials and atmospheric turbulence*,} {\protect\JournalTitle{Journal of the Optical Society of America}} \textbf{66} (1976).

\bibitem{Vedrenne2016}
N.~V{\'e}drenne, J.-M. Conan, C.~Petit \emph{et~al.}, \enquote{Adaptive optics for high data rate satellite to ground laser link,} {\protect\JournalTitle{Proc. SPIE, Free-Space Laser Communication and Atmospheric Propagation XXVIII}} \textbf{9739} (2016).

\bibitem{canuet_statistical_2018}
L.~Canuet, N.~V{\'e}drenne, J.-M. Conan \emph{et~al.}, \enquote{Statistical properties of single-mode fiber coupling of satellite-to-ground laser links partially corrected by adaptive optics,} {\protect\JournalTitle{Journal of the Optical Society of America A}} \textbf{35} (2018).

\bibitem{ma2007quantum}
X.~Ma, C.-H.~F. Fung, and H.-K. Lo, \enquote{Quantum key distribution with entangled photon sources,} {\protect\JournalTitle{Physical Review A}} \textbf{76} (2007).

\bibitem{PhysRevApplied.16.014067}
M.~T. Gruneisen, M.~L. Eickhoff, S.~C. Newey \emph{et~al.}, \enquote{Adaptive-{{Optics-Enabled Quantum Communication}}: {{A Technique}} for {{Daytime Space-To-Earth Links}},} {\protect\JournalTitle{Physical Review Applied}} \textbf{16} (2021).

\bibitem{Tomamichel2012}
M.~Tomamichel, C.~C.~W. Lim, N.~Gisin \emph{et~al.}, \enquote{Tight finite-key analysis for quantum cryptography,} {\protect\JournalTitle{Nature Communications}} \textbf{3} (2012).

\bibitem{PhysRevApplied.18.024059}
W.~Wen, Z.~Chen, L.~Lu \emph{et~al.}, \enquote{Realizing an {{Entanglement-Based Multiuser Quantum Network}} with {{Integrated Photonics}},} {\protect\JournalTitle{Physical Review Applied}} \textbf{18} (2022).

\bibitem{Kneizys1988}
F.~X. Kneizys, E.~P. Shettle, L.~W. Abreu \emph{et~al.}, \emph{Users Guide to LOWTRAN 7}, Air Force Geophysics Laboratory (1988).

\bibitem{modtran}
A.~Berk, P.~Conforti, R.~Kennett \emph{et~al.}, \enquote{{{MODTRAN6}}: A major upgrade of the {{MODTRAN}} radiative transfer code,} {\protect\JournalTitle{Proc. SPIE, Algorithms and Technologies for Multispectral, Hyperspectral, and Ultraspectral Imagery XX}} \textbf{9088} (2014).

\end{thebibliography}

\bibliographyfullrefs{main}


\end{document}